\documentclass[pra,floatfix,showpacs,superscriptaddress]{revtex4}
\usepackage{amssymb}
\usepackage{amsmath}
\usepackage{graphicx}
\usepackage{float}
\usepackage{dcolumn}
\usepackage{bm}

\begin{document}

\title{On the Harmonic approximation for large Josephson junction coupling
charge qubits }
\author{T. Shi}
\affiliation{Department of Physics, Nankai University, Tianjin
300071, China}
\author{B. Chen}
\affiliation{Department of Physics, Nankai University, Tianjin
300071, China}
\author{Z. Song}
\affiliation{Department of Physics, Nankai University, Tianjin
300071, China}
\author{C. P. Sun}\email{suncp@itp.ac.cn; http://www.itp.ac.cn/~suncp}
\affiliation{Department of Physics, Nankai University, Tianjin
300071, China} \affiliation{Institute of Theoretical Physics, The
Chinese Academy of Science, Beijing, 100080, China}


\begin{abstract}
We revisit the harmonic approximation (HA) for a large Josephson junction
interacting with some charge qubits through the variational approach for the
quantum dynamics of the junction-qubit coupling system. By making use of
numerical calculation and analytical treatment, the conditions under which
HA works well can be precisely presented to control the parameters
implementing the two-qubit quantum logical gate through the couplings to the
large junction with harmonic oscillator (HO) Hamiltonian.
\end{abstract}

\maketitle


\section{Introduction}

For quantum information science and technologies, it is crucial to build the
fundamental quantum logic gates \cite{1}. Together with the basic single bit
logic gates, the non-trivial two bit gates constitutes the fundamental
blocks for the quantum network of quantum computing \cite{2}. Physically, the
two-bit quantum logic gate are based on the entanglement of two-qubit
through the controllable interaction between them \cite{3}.

In most scenarios to build the two-qubit logic gates, two qubits interact
with a common object as data bus. When the normal frequency of the data bus
is off resonance with respect to the energy level spacing of the qubits, the
variables of the data bus can be removed and then induce an effective
interaction between the two qubits, which can create an entanglement based
on two-qubit logic gate. Most recently, this idea was used to construct the
two-qubit logic gates of the superconductor Josephson junction where the
data bus is implemented as a large junction. Since the recent experiments \cite{4} have
shown longer time quantum coherence, the Josephson junction qubits ( JJ
qubits ) \cite{5} including charge qubits, flux qubit and the single junction qubit,
the practical schemes to implement the two-gate operations of JJ qubit
become very important. We notice that many of these schemes depend on the
approximate approach to treat the large junction as a harmonic oscillation.
We call this approximation the harmonic approximation (HA). Due to the
linearity of its coupling to the two charge qubits, the variable of the
harmonic oscillator (HO) is easy to remove in a dynamic way or
adiabatically \cite{6}.

Now the validity of HA becomes the focus in search of the implementation of
the JJ qubit based quantum computation. In this paper, we will tackle this
problem by the numerical calculation and the analytical considerations.

\section{The model for the coupling of a large junction to a charge qubit}

We consider a system of coupled charge-phase qubits \cite{7,8} as illustrated by the
electronic circuit shown in Fig.1. We are interested in the limit case that
the large junction (of capacitance $C$) stays at a low excitation or even a
thermal state. The Cooper pair box is a small junction of capacitance $%
C^{^{\prime }}$ ($C^{^{\prime }}=C^{\prime\prime  }$) that form a superconducting loop. $%
C_{g}$ is the capacitance of the gate, and $E_{J}$ is the Josephson coupling
energy. The total Hamiltonian containing the reduced Coulomb energy and the
three Josephson coupling energies \cite{6} reads
\begin{equation}
H=E_{c}(n^{^{\prime }}-n_{g})^{2}-E_{J}^{^{\prime }}(\cos \varphi ^{^{\prime
}}+\cos \varphi ^{\prime\prime })+E_{c}n^{2}-E_{J}\cos \theta
\end{equation}%
where $E_{c}=2e^{2}/C_{\Sigma },$ $n_{g}=C_{g}V_{g}/2e,$ $E_{c}=2e^{2}/C$
and $C_{\Sigma }=C_{g}+C^{^{\prime }}+C^{\prime\prime}$. Here $n^{^{\prime }}$ is the
number of Cooper pair on the island, while $n$ is the number of Cooper pair
on the Coulomb island connected with the large junction. $\varphi ^{^{\prime
}}$, $\varphi ^{\prime\prime}$, and $\theta $ are superconducting-phase differences
across the relevant junction. They are related through the fluxoid
quantization condition around the loop $\theta +\varphi ^{^{\prime }}-$ $%
\varphi ^{\prime\prime}=2\Theta $ and $\Theta _{k}=\pi \Phi _{x}/\phi _{0}$.
Introducting $\varphi =(\varphi ^{^{\prime }}+\varphi ^{\prime\prime})/2$, we rewrite
the Hamiltonian as
\begin{equation}
H=E_{c}(n^{^{\prime }}-n_{g})^{2}+E_{c}n^{2}-E_{J}\cos \theta
-2E_{J}^{^{\prime }}\cos (\frac{\theta }{2}-\Theta )\cos \varphi
\end{equation}%

\vspace*{-1.0cm}
\begin{figure}[h]
\hspace{24pt}\includegraphics[width=8cm,height=12cm]{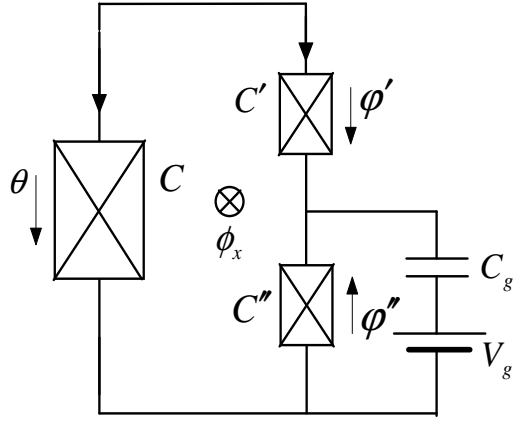}
\vspace*{%
-5.0cm} \caption{Schematic electric circuit diagram of two
charge-phase qubits coupled through a large junction. }
\end{figure}

For charge-phase qubit, when $n_{g}=0.5$ and $E_{J}$ is not much larger than
$E_{c}$, the avove Hamiltonian can be reduced as
\begin{equation}
H=E_{c}(n^{2}-T\cos \theta +T^{^{\prime }}\sin \frac{\theta }{2}\cdot \sigma
_{3})
\end{equation}%
where $T=E_{J}/E_{c},$ $T^{^{\prime }}=E_{J}^{\prime }/E_{c}$ and $\sigma
_{3}$ is Pauli matrix. The operators $n$ and $\theta$ satisfy the
quantization condition $\left[ \theta ,n\right] =i.$

To consider the validity of the HA we only focus on the large junction
together with its coupling to the charge qubit. In this sense we write down
the corresponding Hamiltonian
\begin{eqnarray}
H &=&\hat{n}^{2}-T\cos \theta +T^{\prime }\sin \frac{\theta }{2}\sigma _{3}
\\
&\equiv &\left[
\begin{array}{cc}
H^{(+)} & 0 \\
0 & H^{(-)}%
\end{array}%
\right]   \notag
\end{eqnarray}%
Since the relation $H^{(+)}(\theta )=H^{(-)}(-\theta )$ holds, we know that
these two Hamiltonians $H^{(+)},H^{(-)}$ have the same spectrum. Thus one
can get all the eigenstates of (1) from the eigenstates of $H^{(+)}$. In the
following, we will concentrate on the eigenstates of the Hamiltonian in the
qubits states spanned by discrete charge states $|n>$, $n=0,\pm \frac{1}{2}%
,\pm 1,\pm \frac{3}{2},\pm 2,\cdots ,\pm \infty .$ The periodic potential
can be expanded around the equilibrium points $\theta _{0},$ where $\sin
(\theta _{0}/2)=\pm T^{\prime }/4T$ at which potential has minima. Then $%
H^{(+)}$ can be written as
\begin{equation}
H^{(+)}=\left\{
\begin{array}{c}
\hat{n}^{2}+\frac{16T^{2}-T^{^{\prime }2}}{32T}[\theta -(2\pi i \\
+2\arcsin (-T^{\prime }/4T))]^{2}-\frac{8T^{2}+T^{^{\prime }2}}{8T}, \\
\theta \in \lbrack (2i-1)\pi ,(2i+1)\pi ],i=\text{even} \\
\\
\\
\hat{n}^{2}+\frac{16T^{2}-T^{^{\prime }2}}{32T}[\theta -(2\pi i \\
-2\arcsin (-T^{\prime }/4T))]^{2}-\frac{8T^{2}+T^{^{\prime }2}}{8T} \\
\theta \in \lbrack (2i-1)\pi ,(2i+1)\pi ],i=\text{odd.}%
\end{array}%
\right.
\end{equation}%
In large $T$ limit, the tunnelling between two neighbor wells is forbidden.
The eigen functions of $a_{i}^{\dag }a_{i}$ are standard Harmonic wave
functions with the argument $[\theta -(2\pi i\pm 2\arcsin (-T^{\prime }/4T))]
$ when $i$ is even or odd. Obviously, the spectrum of the Hamiltonian \cite{7} is
the same as the simple harmonic oscillator, i.e.
\begin{eqnarray}
E_{m} &=&(m+\frac{1}{2})\sqrt{\frac{16T^{2}-T^{\prime 2}}{8T}}-T-\frac{%
T^{\prime 2}}{8T} \\
(m &=&0,1,2,......),  \notag
\end{eqnarray}%
but the eigenstates are multi-fold degenerate. Notice that the above
discussion is based on the assumption that there is no tunnelling between
two individual parabolic potential wells. The validity of this assumption
need to be investigated analytically and numerically.

\section{Variational solution for the motion equation}

In this section, we solve the Schrodinger equation governed by large
junction Hamiltonian (2). In the Hilbert space spanned by the eigen state $%
|p>$ of $\widehat{n},\ $where $\widehat{n}|p>=p|p>$ and\ $p\in \lbrack
-\infty ,\infty ]$ the Hamiltonian can be written as
\begin{eqnarray}
H &=&\int_{-\infty }^{\infty }|p><p|p^{2}dp  \notag \\
&&-\frac{T}{2}\int_{-\infty }^{\infty }(|p+1 ><p|+|p><p+1|)dp \\
&&-i\frac{T^{\prime }}{2}\int_{-\infty }^{\infty }(|p+\frac{1}{2}><p|-|p><p+%
\frac{1}{2}|)dp.  \notag
\end{eqnarray}%
Based on the transformation $p\rightarrow n+k$, $\int_{-\infty }^{\infty
}dp\rightarrow \sum_{n}\int_{0}^{1/2}dk$, where $n$ is integer and
half-integer, this Hamiltonian is written as
\begin{eqnarray}
H &=&\frac{1}{2}\int_{0}^{1/2}H^{k}dk  \notag \\
&=&\frac{1}{2}\sum_{n}\int_{0}^{1/2}[|n+k><n+k|(n+k)^{2}  \notag \\
&&-\frac{T}{2}(|n+k+1 ><n+k| \\
&&+|n+k ><n+k+1|)  \notag \\
&&-i\frac{T^{\prime }}{2}(|n+k+\frac{1}{2}><n+k|  \notag \\
&&-|n+k ><n+k+\frac{1}{2}|)]dk.  \notag
\end{eqnarray}%
The Hamiltonian is invariant under the transformation $k\rightarrow N/2+k$,
where $N$ is integer. It indicates that the whole space can be decomposed
into the invariant subspaces denoted by $k$. Therefore the eigenstates of
(4) can be written as
\begin{equation}
|\psi >=\prod_{0\leq k<1/2}|\psi ^{k}>
\end{equation}%
This means that one can obtain the eigenstates in any subspace from $|\psi >$
by the projection operator $P_{k}=\sum_{n}|n+k><n+k|$, whenever $P_{k}|\psi
>\neq 0$. What we concern is the eigenstates in $k=0$ subspace. Since it is
not so easy to tackle the eigen problem analytically in a subspace spanned
by discrete basis, our strategy is that we first study the eigenstate of (3)
by variational approach, then project the eigenstate onto the $k=0$ subspace.

In order to investigate the eigenstates of the Hamiltonian in large $%
T^{\prime }$ limit, we introduce a set of trial harmonic wave
function,
\begin{equation}
|\varphi _{n}>=N_{n}\int_{-\infty }^{\infty }\exp (-\frac{1}{2}\alpha
^{2}p^{2}+i\beta p)H_{n}(\alpha p)dp\left\vert p\right\rangle
\end{equation}%
where $\alpha ,\beta $ are undetermined constants, $N_{n}=\sqrt{\frac{\alpha
}{\sqrt{\pi }2^{n}n!}}$ is renormalization factor. We assume $\alpha \ll 1$.
A direct calculation yields the following matrix elements
\begin{eqnarray}
\left\langle \varphi _{m}\right\vert H|\varphi _{n} &>&=A\delta _{m,n}+B(%
\sqrt{m(m+1)}\delta _{m,n+2}+\sqrt{n(n+1)}\delta _{m,n-2}) \\
&&+iC(m\alpha \frac{N_{m}}{N_{n}}\delta _{m,n+1}-n\alpha \frac{N_{n}}{N_{m}}%
\delta _{m,n-1})  \notag
\end{eqnarray}%
where
\begin{eqnarray}
A &=&\frac{2m+1}{2\alpha ^{2}}+(\frac{m\alpha ^{2}T}{2}-\cos \beta )\exp (-%
\frac{1}{4}\alpha ^{2})  \notag \\
&&+(\frac{m\alpha ^{2}}{8}-1)T^{\prime }\sin \frac{\beta }{2}\exp (-\frac{1}{%
16}\alpha ^{2}) \\
B &=&\frac{1}{2\alpha ^{2}}-\frac{\alpha ^{2}}{16}[4T\cos \beta \exp (-\frac{%
1}{4}\alpha ^{2}) \\
&&+T^{\prime }\sin \frac{\beta }{2}\exp (-\frac{1}{16}\alpha ^{2})] \\
C &=&T\sin \beta \exp (-\frac{1}{4}\alpha ^{2})+\frac{T^{\prime }}{2}\cos
\frac{\beta }{2}\exp (-\frac{1}{16}\alpha ^{2})  \notag
\end{eqnarray}%
The trial wave functions are approximate wave functions of the Hamiltonian
if all the off-diagonal elements vanish. Taking $B=C=0$, we have
\begin{equation}
\sin \frac{\beta }{2}=\frac{T^{\prime }}{4T},\ \ \alpha ^{4}=\frac{32T}{%
16T^{2}-T^{\prime 2}}.
\end{equation}%
The spectrum is
\begin{equation}
E_{m}=(m+\frac{1}{2})\sqrt{\frac{16T^{2}-T^{\prime 2}}{8T}}-T-\frac{%
T^{\prime 2}}{8T}.
\end{equation}%
What we concern is the eigenstate in the subspace $k=0$, which can be
obtained by the project operator $P_{0}=\sum_{n}|n><n|$
\begin{eqnarray}
&&\sum_{n}|n><n|\varphi _{m}>  \notag \\
&=&N_{m}\sum_{n}\exp (-\frac{1}{2}n^{2}\alpha ^{2}+i\beta n)H_{m}(\alpha
n)\left\vert n\right\rangle
\end{eqnarray}%
where $\beta =2\arcsin (\frac{T^{\prime }}{4T})$ and $2\pi -2\arcsin (\frac{%
T^{\prime }}{4T})$ which correspond to the minima of the potential $T\cos
\theta +T^{\prime }\sin \frac{\theta }{2}$. Then in $|n>$ representation,
the lower approximate eigen wave functions are%
\begin{eqnarray}
\psi _{m}^{L}(n) &=&N_{m}\exp (-\frac{1}{2}\alpha ^{\prime
2}n^{2})H_{m}(\alpha ^{\prime }n)\exp (in\theta _{0})  \notag \\
\psi _{m}^{R}(n) &=&N_{m}\exp (-\frac{1}{2}\alpha ^{\prime
2}n^{2})H_{m}(\alpha ^{\prime }n)\exp [in(2\pi -\theta _{0})] \\
\theta _{0} &=&2\arcsin (\frac{T^{\prime }}{4T})  \notag \\
m &=&0,1,2,\cdots ,n=0,\pm \frac{1}{2},\pm 1,\pm \frac{3}{2},\pm 2,\cdots
,\pm \infty .  \notag
\end{eqnarray}%
Notice that each energy level in (16) corresponds to a narrow energy band for
the exact spectrum of the Hamiltonian (2). Employing the Hellmann-Feynman
theorem, one can evaluate the band width
\begin{equation}
\Delta (m)=\int_{0}^{1/2}\langle \frac{\partial H}{\partial k}\rangle _{m}dk=%
\frac{1}{4}
\end{equation}%
which is much less than the gap between two neighbor bands. The eigenstates
of $H^{k}$ share only one level from each band and the approximate wave
function $|\varphi _{n}>$ is just the linear combination of all the
eigenstates corresponding to $E_{n}$.

\section{Numerical results}

In order to verify the validity of HA, numerical calculation is employed to
compute the energy levels and wave functions. Since the whole Hilbert space
is constitutive of invariant subspaces and in the practice system the charge

\vspace*{-0.0cm}
\begin{figure}[h]
\hspace{24pt}\includegraphics[width=10cm,height=15cm]{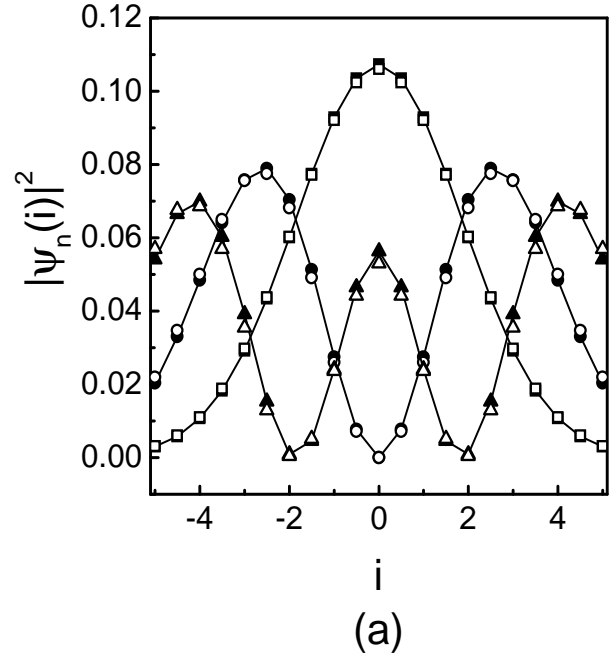}
\vspace*{-5.0cm}

\hspace{24pt}\includegraphics[width=10cm,height=15cm]{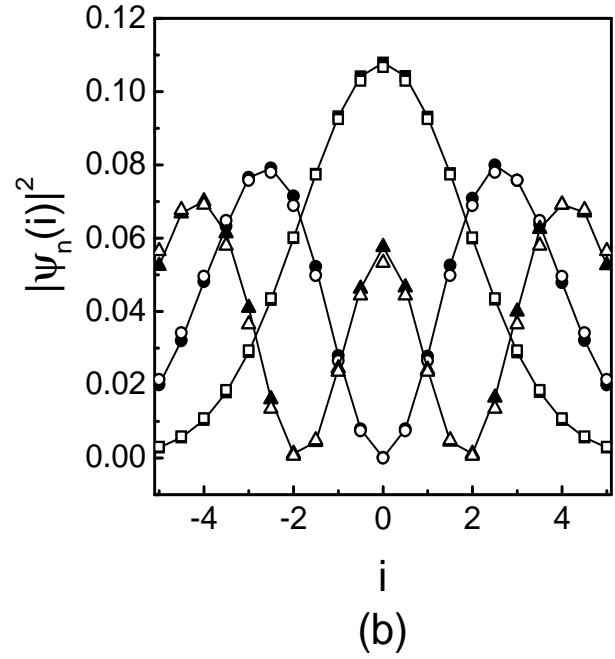}

\vspace*{-4.0cm} \caption{The probability amplitudes of three lowest levels
for $T=100$, (a) $T^{\prime}=0$ and (b) $T^{\prime}=60$ obtained by numerical
and variational methods. Solid ( open ) square, circle and triangle denote numerical
( variational ) results of eigenstates of $n=0, 1, 2$. }
\end{figure}

\vspace*{-5.0cm}
\begin{figure}[h]
\hspace{24pt}\includegraphics[width=10cm,height=15cm]{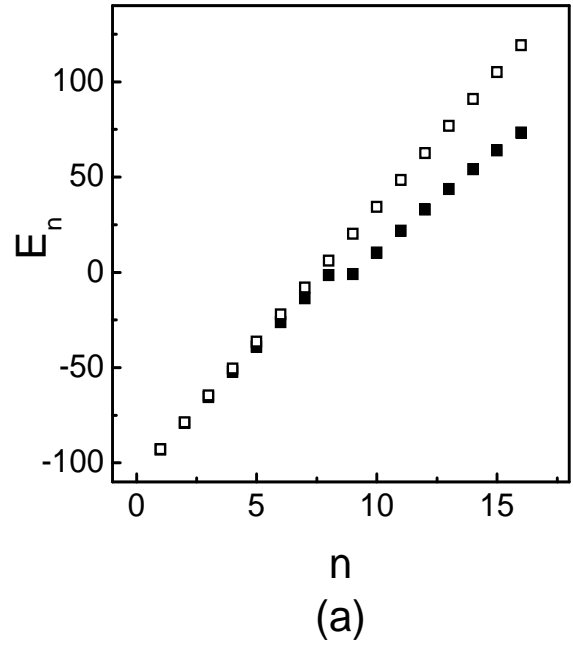}
\vspace*{-5.0cm}

\hspace{24pt}\includegraphics[width=10cm,height=15cm]{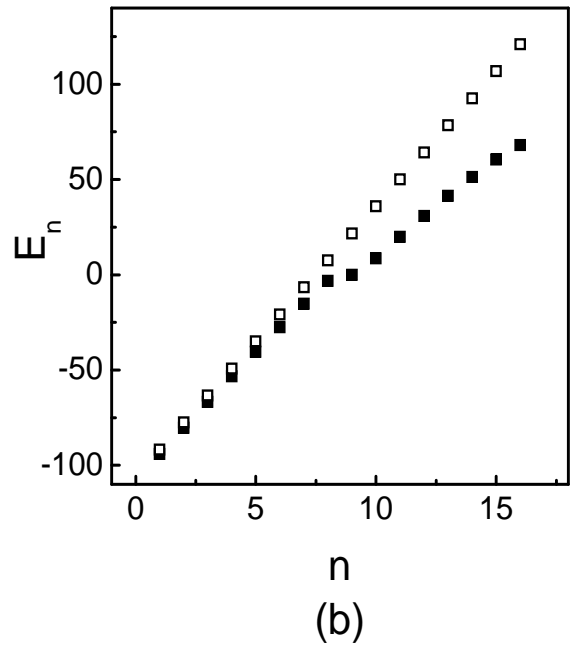}

\vspace*{%
-3.0cm} \caption{The energy levels for
$T=100$, (a) $T^{\prime}=0$ and (b) $T^{\prime}=60$
obtained by numerical ( solid square ) and variational ( open
square ) methods. It shows that the value of $T'$ doesn't effect
the results obtained by two methods so much which agree quite
well for lower levels. }
\end{figure}

\newpage
number $n$ is restricted to be integer and half-integer, exact
diagonalization is performed to get the lower energy spectrum and
corresponding eigen functions of the Hamiltonian in $k=0$ subspace. In Fig.2
and 3, the energy levels for systems with $T=100, T^{\prime }=0$ and $%
T=100, T^{\prime }=60$ obtained by numerical and variational methods are
plotted. It shows that the HA is valid for wide range of $T^{\prime }$. The
sudden deviations occur around zero energy since the tunnelling effect
between neighbor wells is non-neglectable. Under the threshold point
numerical and variational results agree well. In Fig.4 and 5, the
probability amplitudes of three lowest levels for systems with $%
T=100, T^{\prime }=0$ and $T=100, T^{\prime }=60$ obtained by numerical and
variational methods are plotted. It shows that numerical and variational
results agree well.

\section*{Acknowledgments}

This work is supported by the Cooperation Foundation of Nankai and Tianjin
university for research of nanoscience.



\end{document}